\documentclass[showpacs,preprintnumbers,amsmath,amssymb,APSl,prd,nofootinbib,superscriptaddress, twocolumn]{revtex4-2}

\usepackage{amsmath}
\usepackage{amssymb}
\usepackage{amsfonts}
\usepackage{graphicx,bm}
\usepackage{dcolumn}
\usepackage[colorlinks=true]{hyperref}
\usepackage{epsf}
\usepackage{enumerate}
\usepackage{hhline}
\usepackage{array}
\usepackage{tabularx}
\usepackage{color}
\usepackage{subfigure}

\newcommand{\be}{\begin{equation}}
\newcommand{\ee}{\end{equation}}
\newcommand{\bea}{\begin{eqnarray}}
\newcommand{\eea}{\end{eqnarray}}
\newcommand{\beaa}{\begin{eqnarray*}}
\newcommand{\eeaa}{\end{eqnarray*}}

\def\be{\begin{equation}}
\def\ee{\end{equation}}
\def\bea{\begin{eqnarray}}
\def\eea{\end{eqnarray}}

\begin{document}
\title{Propagation and emission of gravitational waves in the weak-field limit within the Palatini formalism}

\author{Albert Duran-Cabac\'es} \email{albert.duran22@uva.es}
\affiliation{Department of Theoretical Physics, Atomic and Optics, Campus Miguel Delibes, \\ University of Valladolid UVA, Paseo Bel\'en, 7,
47011 - Valladolid, Spain}

\author{Diego S\'aez-Chill\'on G\'omez}
\email{diego.saez@uva.es} 
\affiliation{Department of Theoretical Physics, Atomic and Optics, Campus Miguel Delibes, \\ University of Valladolid UVA, Paseo Bel\'en, 7,
47011 - Valladolid, Spain}
\affiliation{Department of Physics, Universidade Federal do Cear\'a (UFC), Campus do Pici, Fortaleza - CE, C.P. 6030, 60455-760 - Brazil}

\begin{abstract}
In the era of gravitational waves physics, when detections of wave fronts are increasing in number, sensitivity, frequencies and distances, gravitational physics has entered a period of maximum activity and brilliance. This has open a new window where General Relativity can be challenged in both weak as strong-field regimes. In this paper, we focus on the analysis of gravitational waves propagation and emission in the weak-field regime for gravitational theories within the Palatini formalism. Our results show that gravitational waves propagation in vacuum matches General Relativity predictions as well as the functional form of the multipolar expansion when considering weak sources. However, a rescaling of the gravitational constant arises, which affects the energy radiated by the gravitational waves emission.

\end{abstract}
%
%
%\pacs{04.50.Kd, 98.80.-k, 95.36.+x}
%%
%%
\maketitle
%
%\def\thesection{\Roman{section}}
%\def\theequation{\Roman{section}.\arabic{equation}}
%
%
%%%%%%%%%%%%%%%%%%%%%%%%%%%%%%%%%%%%%
\section{Introduction}
%%%%%%%%%%%%%%%%%%%%%%%%%%%%%%%%%%%%%

Over the last years, the field of gravitational waves (GWs) has experienced a great growth, mainly pushed forward by the many detections of GWs fronts by the interferometers of LIGO/VIRGO 
(for some of the first detections, see Refs.~\cite{LIGOScientific:2016aoc}).  Such detections were the product of the coalescence of binary black hole systems as well as neutron stars, that allow also to detect the electromagnetic counterpart \cite{LIGOScientific:2017zic,LIGOScientific:2017ync}. Recently, the collaboration KAGRA has joined to network, which will increase the distance of the events and the number of coalescence binary neutron stars systems. In the next incoming years, other detectors such as LISA, Taiji, TianQin or later the Einstein Telescope will start operations where the sensitivity and the range of frequencies will be increased allowing to detect sources at much higher redshifts and also GWs' emission during the inspiral phase of some binary systems (see Refs.~\cite{LISA:2022kgy,Maggiore:2019uih}). Hence, one can say that one of the areas where gravitational physics will be focused  for the next decades is GWs physics. From the theoretical point of view, this is a great opportunity to test the limits of General Relativity (GR). As has been widely analysed, GR has been an incredible successful theory despite also some limitations have shown up. From cosmology to compact objects physics, GR lacks of providing some convince explanations. One may cite dark energy, black holes singularities or the absence of ultraviolet completion of the theory, among others. Hence, along the next years GR will be tested as never has been before and compact objects physics will play an essential role, mainly through the analysis of GWs emissions \cite{Cardoso:2019rvt}. Moreover, the incoming observational data at the cosmological level will also provide another independent source to test GR \cite{Ishak:2018his}. \\

Over the last decades, modifications/extensions of GR have been widely studied, motivated by different phenomena, such analysis has led to a huge increase of knowledge on the way GR might be modified successfully (for a review see Refs.~\cite{Nojiri:2017ncd,Clifton:2011jh,CANTATA:2021ktz}). One of the ways of modifying GR that has been widely explored in the literature is to consider the connection as an independent field \cite{Olmo:2011uz,BeltranJimenez:2017doy,Vitagliano:2010sr}. Despite GR is constructed as a geometrical theory that describes pseudo-Riemannian manifolds that are completely characterised by the spacetime metric, none of its principles states that the connection must be the Levi-Civita connection, i.e. that the metricity condition must hold. In addition, GR might be formulated in two different and physically equivalent ways, where the central magnitude is not the curvature but the torsion or the non-metricity \cite{BeltranJimenez:2019esp}. For such theories, known as Teleparallel gravity and Coincident General Relativity respectively, the connection is not given by the Levi-Civita connection, but it is neither let as a free field. However, this has shown that GR might be modified in very different ways. In this paper, we focus on a particular case of the so-called Ricci Based Gravity (RBG) theories, whose starting point is a gravitational action that depends solely on contractions and products of the Ricci tensor, which is constructed with an a priori independent connection \cite{BeltranJimenez:2017doy,Vitagliano:2010sr}. By following the Palatini formalism, which consists on varying the action with respect to the connection as an independent field, the corresponding field equations are obtained for the connection. As far as the gravitational action turns out the Hilbert-Einstein action, the connection becomes the Levi-Civita connection of GR \cite{Ferraris:1982wci}. One of the main advantages of RBG theories is that the field equations for the spacetime metric remain second order while the equations for the connection leads to a direct solution that establishes a connection compatible to a metric tensor. Then, a particular mapping has been established that leads to the same equations as in GR, but now in terms of other metric tensor related to the spacetime metric a conformal transformation \cite{Afonso:2018bpv}.The simplest version of RBG theories is the so-called Palatini $f(R)$ gravity where the action is given by a function of the contraction of the Ricci tensor scalar. The main advantage of Palatini $f(R)$ gravity is that the field equations for the spacetime metric remain second order while the equations for the connection leads to a direct solution that establishes a connection compatible to a metric tensor. These theories have been widely analysed in different frameworks, particularly a big effort has been followed for analysing the cosmological implications within these theories, as cosmological late-time acceleration \cite{Baghram:2009we,Aoki:2018lwx,Leanizbarrutia:2017xyd,Rosa:2017jld,Rosa:2021ish} or inflationary models \cite{Shimada:2018lnm,Gialamas:2020snr,Antoniadis:2018ywb,Rasanen:2018ihz,Bekov:2020dww,Karam:2021sno,Dioguardi:2021fmr,Dioguardi:2023jwa}. Moreover, an intensive study has drawn a lot of attention in solutions of compact objects, since one can construct solutions for regular black holes/wormholes \cite{Olmo:2015axa,Guerrero:2020uhn,Rosa:2020uoi,Odintsov:2014yaa,Gomez:2020rnq} or describe well stellar interiors \cite{Olmo:2019flu,Olmo:2020fri}.  \\

The present paper intends to focus on the analysis of GWs propagation and emission in the weak-field regime within Palatini $f(R)$ gravity. In general, scalar-tensor theories such as metric $f(R)$ gravities, lead to additional propagating scalar modes that yield to an additional longitudinal polarisation \cite{Capozziello:2008rq,Geng:2012zc,Nojiri:2003ft,Berry:2011pb,Naf:2011za,Barraco:2000dx,Liang:2017ahj,Gogoi:2019zaz}. However, some modifications of GR, such as the Palatini formalism itself  \cite{Lu:2020eux,Bora:2022qwe,Sharif:2014zra} or some extensions of Teleparallel gravity \cite{Bamba:2013ooa} do not contain extra propagating modes. So far, GWs properties in certain theories of modified gravities have been found by using the post-Newtonian formalism \cite{Battista:2021rlh,Battista:2022hmv,DeFalco:2023djo,DeFalco:2024ojf} and the WKB approach \cite{Harikumar:2023gzh}. The analysis of GWs in modified gravities is essential to test the limits of GR and to find out the way GR might be extended. In this sense, some tests have been suggested in the literature, specially for metric $f(R)$ gravities where constraints on the free parameters of the models can be obtained \cite{Vainio:2016qas,Capozziello:2008fn}. Moreover, some modifications of GR might affect stochastic GWs backgrounds, specially during early cosmology \cite{Ananda:2007xh,Odintsov:2021kup,Papanikolaou:2021uhe}. In order to focus our analysis on the study of GWs in the weak-field limit for Palatini $f(R)$ gravity, we express this theory as a scalar-tensor theory (Brans-Dicke-like model). GWs have been widely explored within Brans-Dicke theories, both in the strong-field regime \cite{Kwon:1986dw} as in the weak-field limit (see Refs.~\cite{Ozer:2021qjb,Liu:2022qcx,Du:2018txo,Ohashi:1996uz,Barros:1997qt}). Nevertheless, contrary to usual Brans-Dicke-like theories, the scalar field is not dynamical in Palatini $f(R)$ gravities. Then, by following such procedure we show that propagating GWs in vacuum coincide with GR predictions, as shown previously in \cite{Harikumar:2023gzh} by following a different approach.  Moreover, by considering weak sources, the corresponding multipole expansion is performed, where the quadrupole and octupole moments are computed, showing that despite the GWs equation is sourced by an effective energy-momentum tensor that includes derivatives of the trace of the energy-momentum tensor, the functional form for each order of the multipole expansion reduce to the ones predicted by GR. However, every moment in the multipole expansion is rescaled by the expected value of the scalar field in vacuum, such that the energy radiated by the GWs emission is rescaled accordingly. This might open a window to constrain this type of theories. \\

The paper is organised as follows: modified Palatini $f(R)$ gravity is introduced in section \ref{background}. Section \ref{linearisation} is devoted to the linearisation of the equations. Then, the propagation GWs equation in vacuum is obtained in section \ref{Vacuum}. Whereas the emission of GWs in the weak-field limit is analysed in section \ref{Emission}. Finally, the last section \ref{conclusions} gathers the conclusions of the paper.

%%%%%%%%%%%%%%%%%%%%%%%%%%%%%%%%%%%%%
\section{Modified Palatini $f(R)$ gravity}
\label{background}
%%%%%%%%%%%%%%%%%%%%%%%%%%%%%%%%%%%%%
Along this paper we focus on the so-called Palatini $f(R)$ gravity, where the spacetime metric and the connection are considered as independent fields. The gravitational action is given by:
\be
S=\frac{1}{2\kappa^2}\int dx^4 \sqrt{-g} \left[f(\mathcal{R}) +L_m\right] \ ,
\label{fRaction}
\ee
where $L_m$ is the Lagrangian that describes the matter fields and does not depend on the connection. The Ricci tensor is provided in terms of the independent connection as follows:
\be
\mathcal{R}_{\mu\nu}(\Gamma)=\partial_{\lambda}\Gamma^{\lambda}_{\mu\nu}-\partial_{\nu}\Gamma^{\lambda}_{\mu\lambda}+\Gamma^{\lambda}_{\sigma\lambda}\Gamma^{\sigma}_{\mu\nu}-\Gamma^{\lambda}_{\sigma\nu}\Gamma^{\sigma}_{\mu\lambda}\ . 
\label{Ricci}
\ee
Whereas the curvature scalar for the gravitational action is $\mathcal{R}=g^{\mu\nu}\mathcal{R}_{\mu\nu}(\Gamma)$. The independent connection is not necessarily torsionless a priori, however, only its symmetric part plays a role on the equations of motion (see. \cite{Olmo:2011uz}). Note that for the Hilbert-Einstein action, the Palatini formalism directly leads to GR, since the metricity condition is recovered, as shown below. However, for a non-linear function of $\mathcal{R}$ in (\ref{fRaction}), the metricity condition does not hold and the connection remains independent in principle. Then, by varying the action (\ref{fRaction}) with respect to the spacetime metric, the field equations are obtained \cite{Olmo:2011uz}:
\be
  f_{\mathcal{R}}\mathcal{R}_{\mu\nu}-\frac{1}{2}g_{\mu\nu}f=\kappa^2 T_{\mu\nu}\ ,
  \label{Fieldeqs1}
  \ee
where $T_{\mu\nu}=-\frac{2}{\sqrt{-g}}\frac{\delta  (\sqrt{-g} L_m)}{\delta g^{\mu\nu}}$ is the energy-momentum tensor and  $f_{\mathcal{R}}=\frac{df}{d\mathcal{R}}$. In addition, variations of the Ricci tensor with respect to the connection in the action lead to:
\be
  \tilde{\nabla}_{\lambda}\left(\sqrt{-g}f_{\mathcal{R}}g^{\mu\nu}\right)=0\ .
  \label{Fieldeqs}
  \ee
Here $\tilde{\nabla}$ is the covariant derivative defined by the connection $\Gamma$. This equation gives the solution for the connection $\Gamma$ in terms of the Levi-Civita connection for the metric tensor:
\be
q_{\mu\nu}=\Omega^2 g_{\mu\nu}\ , \quad \Omega^2=f_{\mathcal{R}}\ ,
\label{conform_transform22}
\ee
Hence, one has $\tilde{\nabla}_{\lambda}\left(\sqrt{-q}q^{\mu\nu}\right)=0$, which reproduces the metricity condition for the covariant derivative $\tilde{\nabla}$ for the metric $q_{\mu\nu}$.  Moreover, by taking the trace of the field equations (\ref{Fieldeqs1}), an algebraic relation between the scalar $\mathcal{R}$ and the energy-momentum tensor $T$ is obtained:
\be
  f_{\mathcal{R}}\mathcal{R}-2f=\kappa^2 T\ \quad \rightarrow \quad \mathcal{R}=\mathcal{R}(T)\ .
   \label{traceRT}
   \ee
By using the conformal relation given in (\ref{conform_transform22}), the field equations (\ref{Fieldeqs1}) may be rewritten in terms of the spacetime metric, instead of the independent connection. To do so, one has to make a conformal transformation of the Ricci tensor as follows:
\begin{align}
\mathcal{R}_{\mu\nu}(q)=R_{\mu\nu}(g)+\frac{4}{\Omega^2}\nabla_{\mu}\Omega\nabla_{\nu}\Omega-\frac{2}{\Omega}\nabla_{\mu}\nabla_{\nu}\Omega \nonumber \\
-g_{\mu\nu}\frac{g^{\rho\sigma}}{\Omega^2}\nabla_{\rho}\Omega\nabla_{\sigma}\Omega-g_{\mu\nu}\frac{\Box\Omega}{\Omega}\ .
\label{ConformalR}
\end{align}

Now the covariant derivatives shown in the r.h.s. of (\ref{ConformalR}) are compatible with the spacetime metric $g_{\mu\nu}$. Hence, the field equations (\ref{Fieldeqs1}) can be expressed as:
\begin{align}
R_{\mu\nu}(g)-\frac{1}{2}g_{\mu\nu}R(g)=\frac{\kappa^2}{f_{\mathcal{R}}}T_{\mu\nu}-g_{\mu\nu}\frac{\mathcal{R}f_{\mathcal{R}}-f}{2f_{\mathcal{R}}} \nonumber \\
-\frac{3}{2f_{\mathcal{R}}^2}\left[\nabla_{\mu}f_{\mathcal{R}}\nabla_{\nu}f_{\mathcal{R}}-\frac{1}{2}g_{\nu\mu}\nabla_{\lambda}f_{\mathcal{R}}\nabla^{\lambda}f_{\mathcal{R}}\right] \nonumber \\+\frac{1}{f_{\mathcal{R}}}\left[\nabla_{\mu}\nabla_{\nu}f_{\mathcal{R}}-g_{\mu\nu}\Box f_{\mathcal{R}}\right]\ .
\label{fieldEq2}
\end{align}
In addition, it is straightforward to show that the field equations (\ref{fieldEq2}) are equivalent to those of a scalar-tensor theory by defining the following scalar field $\phi$: 
\begin{equation}
    \phi=f_{\mathcal{R}}, \quad \quad \quad V(\phi)=\mathcal{R}\phi-f(\mathcal{R})\ .
\end{equation}
Then, the field equations yield:
\begin{align}
    R_{\mu \nu}-\frac{1}{2} g_{\mu \nu} R = \frac{\kappa^2}{\phi}T_{\mu \nu}-\frac{V(\phi)}{2\phi}g_{\mu\nu} \nonumber \\
    -\frac{3}{2\phi^2} \left( \partial_\mu \phi \partial_\nu \phi- \frac{1}{2} g_{\mu \nu}(\partial \phi)^2 \right) \nonumber \\
    + \frac{1}{\phi}\left( \nabla_\mu \nabla_\nu \phi-g_{\mu \nu} \Box \phi \right)\ ,
    \label{fieldBD}
\end{align}
Note also that the trace equation (\ref{traceRT}) can be written also in terms of the scalar field as:
\begin{equation}
    2V(\phi)-\phi V'(\phi) = \kappa^2 T\ ,
    \label{potentialBD}
\end{equation}
where in the last equation we have used the fact that $\mathcal{R}=V'(\phi)$. Note that contrary to the usual Brans-Dicke theories, here the scalar field equation $(\ref{potentialBD})$ is an algebraic equation which does not involve derivatives of the scalar field. Equations (\ref{fieldBD}-\ref{potentialBD}) are the starting point of our analysis. In the next section, the equations are linearised by considering a small perturbation on a background arbitrary metric.

%%%%%%%%%%%%%%%%%%%%%%%%%%%%%%%%%%%
\section{Linearisation of the field equations}
\label{linearisation}

%%%%%%%%%%%%%%%%%%%%%%%%%%%%%%%%%%
Let us consider a small perturbation on the metric, on the scalar field and on the matter energy-momentum tensor around a particular background solution of the field equations:
\begin{equation}
    g_{\mu \nu}= \overline{g}_{\mu \nu} + h_{\mu \nu}, \quad T_{\mu \nu}=\overline{T}_{\mu \nu}+\delta T_{\mu \nu}, \quad \phi=\overline{\phi}+\delta \phi\ ,
    \label{perturbations1}
\end{equation}
where the over-line is used to denote the background quantities, while $\delta T_{\mu \nu}$ and $\delta \phi$ are of the same order than the perturbation $h_{\mu \nu}$. The Einstein tensor at the l.h.s. of Eq. \eqref{fieldBD} can be expanded in the following standard way:
\begin{align}
    G_{\mu\nu}&=R_{\mu \nu}-\frac{1}{2} g_{\mu \nu} R \nonumber \\
    &=\overline{G}_{\mu \nu}+\frac{1}{2} \big( \nabla^\lambda \nabla_\mu h_{\lambda \nu} + \nabla^\lambda \nabla_\nu h_{\lambda \mu}-\Box h_{\mu \nu} \nonumber \\
    &- \nabla_\nu \nabla_\mu h + \Box h \overline{g}_{\mu \nu}- \nabla_\alpha \nabla_\beta h^{\alpha \beta} \overline{g}_{\mu \nu}-\overline{R} h_{\mu \nu} \nonumber \\
    &+\overline{g}_{\mu \nu} h^{\alpha \beta}\overline{R}_{\alpha \beta} \big),
\end{align}
where $h=g^{\mu \nu}h_{\mu \nu}$, which at first order reads $h=\overline{g}^{\mu \nu}h_{\mu \nu}$. The covariant derivatives and D'Alembertians are also computed  by using the background metric. Whereas the r.h.s. of field Eqs. \eqref{fieldBD} at first order in perturbations lead to:
\begin{widetext}
\begin{align}
    \kappa^2 \frac{\overline{T}_{\mu \nu}}{\overline{\phi}}+\kappa^2 \frac{\delta T_{\mu \nu }}{\overline{\phi}}-\kappa^2 \frac{\overline{T}_{\mu \nu}}{\overline{\phi}}\frac{\delta \phi}{\overline{\phi}}-\frac{V(\overline{\phi})}{2 \overline{\phi}} \overline{g}_{\mu \nu} - \frac{V(\overline{\phi})}{2 \overline{\phi}} h_{\mu \nu}+\frac{V(\overline{\phi})}{2 \overline{\phi}} \overline{g}_{\mu \nu}\frac{\delta \phi}{\overline{\phi}} -\frac{V'(\overline{\phi})}{2} \overline{g}_{\mu \nu}\frac{\delta \phi}{\overline{\phi}}\nonumber\\
    -\left( \frac{3}{2\overline{\phi}^2}-\frac{3}{\overline{\phi}^2} \frac{\delta \phi}{\overline{\phi}} \right) \left[ \partial_\mu \overline{\phi} \partial_\nu \overline{\phi} +\partial_\mu \overline{\phi} \partial_\nu \delta \phi + \partial_\mu \delta \phi \partial_\nu \overline{\phi}- \frac{1}{2} \overline{g}_{\nu \mu} (\partial \overline{\phi})^2-\frac{1}{2} h_{\mu \nu} (\partial \overline{\phi})^2-\overline{g}_{\mu \nu} \partial^\alpha \partial_\alpha \delta \phi \right] \nonumber \\
    +\frac{1}{\overline{\phi}} \Bigg[ \nabla_\mu \nabla_\nu \overline{\phi} + \nabla_\mu \nabla_\nu \delta\phi-\frac{\overline{g}^{\lambda \alpha}}{2}(\nabla_\nu h_{\alpha \mu}+\nabla_\mu h_{\alpha \nu}-\nabla_\alpha h_{\mu \nu}) \partial_\lambda \overline{\phi}- \overline{g}_{\mu \nu} \Box \overline{\phi}- \overline{g}_{\mu \nu} \Box \delta \phi \nonumber\\
    +\overline{g}_{\mu\nu}\frac{\overline{g}^{\lambda \alpha}}{2} (\nabla_\beta h_{\lambda \alpha}+\nabla_\alpha h_{\lambda \beta}-\nabla_\lambda h_{\alpha \beta})-h_{\mu\nu} \Box \overline{\phi} -\nabla_\mu \nabla_\nu \overline{\phi}\frac{\delta \phi}{\overline{\phi}} - \overline{g}_{\mu \nu} \Box \overline{\phi}\frac{\delta \phi}{\overline{\phi}} \Bigg].
    \end{align}

Hence, by taking both sides, finally the field equations at first order are obtained:

    \begin{align}
   \frac{1}{2} \left( \nabla^\lambda \nabla_\mu h_{\lambda \nu} + \nabla^\lambda \nabla_\nu h_{\lambda \mu}-\Box h_{\mu \nu}- \nabla_\nu \nabla_\mu h + \Box h \overline{g}_{\mu \nu}- \nabla_\alpha \nabla_\beta h^{\alpha \beta} \overline{g}_{\mu \nu}-\overline{R} h_{\mu \nu}+\overline{g}_{\mu \nu} h^{\alpha \beta}\overline{R}_{\alpha \beta} \right)= \nonumber \\ 
    =\kappa^2 \frac{\delta T_{\mu \nu }}{\overline{\phi}}-\kappa^2 \frac{\overline{T}_{\mu \nu}}{\overline{\phi}}\frac{\delta \phi}{\overline{\phi}} - \frac{V(\overline{\phi})}{2 \overline{\phi}} h_{\mu \nu}+\frac{V(\overline{\phi})}{2 \overline{\phi}} \overline{g}_{\mu \nu}\frac{\delta \phi}{\overline{\phi}}-\frac{V'(\overline{\phi})}{2} \overline{g}_{\mu \nu}\frac{\delta \phi}{\overline{\phi}} \nonumber\\
    -\frac{3}{2\overline{\phi}^2} \left[ \partial_\mu \overline{\phi} \partial_\nu \delta \phi + \partial_\mu \delta \phi \partial_\nu \overline{\phi}-\frac{1}{2} h_{\mu \nu} (\partial \overline{\phi})^2-\overline{g}_{\mu \nu} \partial^\alpha \overline{\phi} \partial_\alpha \delta \phi \right] + \frac{3}{\overline{\phi}^2} \frac{\delta \phi}{\overline{\phi}}\left[ \partial_\mu \overline{\phi} \partial_\nu \overline{\phi}- \frac{1}{2} \overline{g}_{\nu \mu} (\partial \overline{\phi})^2 \right]\nonumber \\
    +\frac{1}{\overline{\phi}} \Bigg[ \nabla_\mu \nabla_\nu \delta\phi-\frac{\overline{g}^{\lambda \alpha}}{2}(\nabla_\nu h_{\alpha \mu}+\nabla_\mu h_{\alpha \nu}-\nabla_\alpha h_{\mu \nu}) \partial_\lambda \overline{\phi}- \overline{g}_{\mu \nu} \Box \delta \phi \nonumber\\
    +\overline{g}_{\mu\nu}\frac{\overline{g}^{\lambda \alpha}}{2} (\nabla_\beta h_{\lambda \alpha}+\nabla_\alpha h_{\lambda \beta}-\nabla_\lambda h_{\alpha \beta})\partial_\beta \overline{\phi}-h_{\mu\nu} \Box \overline{\phi} -\nabla_\mu \nabla_\nu \overline{\phi}\frac{\delta \phi}{\overline{\phi}} - \overline{g}_{\mu \nu} \Box \overline{\phi}\frac{\delta \phi}{\overline{\phi}} \Bigg].
    \label{perturbedfield}
    \end{align}
     \end{widetext}
In addition, the scalar field equation at first order \eqref{potentialBD} turns out:
    \begin{equation}
        \left(V'(\overline{\phi})- \overline{\phi} V''(\overline{\phi})\right)\delta \phi=\kappa^2 \delta T.
        \label{perturbedpotential}
    \end{equation}
   
From here, we can now analyse the propagation of gravitational waves in vacuum and also the generation of them in the weak field limit. 

%%%%%%%%%%%%%%%%%%%%%%%%%%%%%%%%%%%%%%
\section{Propagation of gravitational waves in vacuum}
\label{Vacuum}
%%%%%%%%%%%%%%%%%%%%%%%%%%%%%%%%%%%%%%

While the linearisation of the field equations can be used to analyse more complex scenarios, as black hole perturbations and the corresponding stability, here we aim to focus on the simple aspect of the propagation in vacuum of the GW.  Hence, for the weak field limit and by considering the background spacetime metric as the Minkowski metric, one has:
 \be
 \overline{R}=0\ ,\quad \overline{R}_{\alpha \beta}=0 , \quad  \overline{T}_{\mu \nu}=0\ , \quad \overline{\phi}=\phi_0\ .
 \label{vacuumCond}
 \ee
 
Then, by definition, $V'(\phi_0)=0$, and in order to satisfy the background solution $V(\phi_0)=0$ must hold. Hence,  the equation for the perturbations \eqref{perturbedfield} reads:
\begin{align}
    \frac{1}{2} \big( \nabla^\lambda \nabla_\mu h_{\lambda \nu} + \nabla^\lambda \nabla_\nu h_{\lambda \mu}-\Box h_{\mu \nu} \nonumber \\
   - \nabla_\nu \nabla_\mu h \nonumber + \Box h \eta_{\mu \nu}- \nabla_\alpha \nabla_\beta h^{\alpha \beta} \eta_{\mu \nu} \big) \nonumber \\
   =\nabla_\mu \nabla_\nu \frac{\delta\phi}{\phi_0}-\eta_{\mu \nu} \Box \frac{\delta\phi}{\phi_0}\ .
    \label{perturbedvacuum}
\end{align}

This equation is free from gauge choices and covariant derivatives are actually partial derivatives, as correspond for a flat background metric. In order to simplify the equations (\ref{perturbedvacuum}), we can choose the Lorentz Gauge $\partial^\alpha \theta_{\alpha \beta}=0$ to obtain the usual wave equation. Note also that the r.h.s. of the equations (\ref{perturbedvacuum}) automatically satisfies the Lorentz gauge. By defining the new tensor $\theta_{\mu \nu}=h_{\mu \nu}-\frac{1}{2} h \eta_{\mu \nu}$ the gravitational wave equation yields:
\begin{equation}
    \Box \theta_{\mu \nu}=-2\nabla_\mu \nabla_\nu \frac{\delta\phi}{\phi_0}+2\eta_{\mu \nu} \Box \frac{\delta\phi}{\phi_0}\ .
    \label{waveq}
\end{equation}
In comparison to GR, now the GWs equations in vacuum are not effectively in vacuum but are sourced by the perturbation of the scalar field. To show that actually such a perturbation does not play any role (in vacuum), let's see that in vacuum Eq. \eqref{perturbedpotential} implies:
\begin{equation}
       - \phi_0 V''(\phi_0)\delta \phi=0\ ,
\end{equation}
However, this equation may imply three different scenarios depending actually on the theory:
\begin{itemize}
    \item For $\phi_0\neq0$ and $V''(\phi_0)\neq0$, the perturbation for the scalar field becomes null $\delta\phi=0$ and consequently, the GW equation in vacuum turns out the same as in GR.
        \item For $ \phi_0=0$. This would imply that that $\phi_0=f_{\mathcal{R}_{0}}=0$ and consequently the field equations (\ref{fieldEq2}) or (\ref{fieldBD}) are ill-defined for a Minkowski background, which does not correspond to a viable theory. 
    \item $V''(\phi_0)=0$. This is probably the most interesting case, as it might be in principle possible as far as the scalar field potential and consequently the gravitational action $f_{\mathcal{R}}$ satisfies such condition. However, for the vacuum case $T=0$, the non-perturbative equation (\ref{potentialBD}) implies that the scalar field becomes constant, so the perturbation $\delta\phi$ turns out a constant as well, which implies that the GW equation (\ref{waveq}) reduces again to the standard case of GR.
    \end{itemize}
Hence, the only viable possibilities makes the GW equation (\ref{waveq}) to become:
\begin{equation}
    \Box \theta_{\mu \nu}=0\ .
    \label{waveqGR}
\end{equation}
This is the GW equation for GR in vacuum, such that GWs in the Palatini formalism behave the same way in vacuum as in GR, with the same speed of propagation $c$ and the same two types of polarisations. Note that this is not the case for most of scalar-tensor theories, including metric $f(R)$ gravity, where some new longitudinal polarisations arise and/or the waves do not propagate at the speed of light. However, in the next section, we show that even in the weak field limit, when sources provided by an energy-momentum tensor are present, solutions become different as in GR.

%%%%%%%%%%%%%%%%%%%%%%%%%%%%%%%%%%%%%%
\section{Emission of gravitational waves in the weak-field limit}
\label{Emission}
%%%%%%%%%%%%%%%%%%%%%%%%%%%%%%%%%%%%%%

Our aim here is to show the implications for the emission of GWs in Palatini $f(R)$ theories by analysing the field equations when considering weak sources. To do so, we follow the usual approach i.e. we assume the sources to be weak enough in order to consider the background spacetime metric as approximately Minkowskian in vacuum. Then, the energy-momentum tensor can be split into a zero order perturbation and a first order one as given in (\ref{perturbations1}), where the zero order is taken null $\overline{T}_{\mu \nu}\sim0$, such that the energy-momentum tensor arises as a first order perturbation, $T_{\mu\nu}\sim \delta T_{\mu\nu}$ (for more details, see for instance \cite{Maggiore:2007ulw}). By assuming the Lorentz gauge, the equations for the GWs (\ref{perturbedfield}) reduce to:
\begin{equation}
    \Box \theta_{\mu \nu}=-\frac{2\kappa^2}{\phi_0}T_{\mu\nu}-2\nabla_\mu \nabla_\nu \frac{\delta\phi}{\phi_0}+2\eta_{\mu \nu} \Box \frac{\delta\phi}{\phi_0}\ ,
    \label{waveqTmunu}
\end{equation}
where remind that we have defined $\theta_{\mu \nu}=h_{\mu \nu}-\frac{1}{2} h \eta_{\mu \nu}$. While the equation that relates the scalar field perturbation with the energy-momentum tensor (\ref{perturbedpotential}) becomes:
\begin{equation}
       - \phi_0 V''(\phi_0)\delta \phi=\kappa^2 T\ .
       \label{scalarT}
\end{equation}
Note that now the scalar field perturbation depends directly on the source, specifically on the trace of the energy-momentum tensor, such that for traceless matter, as radiation, the perturbation turns out null and the usual GWs equation of GR are recovered again. However, besides such a case, the scalar field perturbation will not be in general null neither constant. In fact, by combining both equations (\ref{waveqTmunu}) and (\ref{scalarT}), the situation becomes clearer:
\begin{align}
    \Box \theta_{\mu \nu}=-\frac{2\kappa^2}{\phi_0}T_{\mu\nu} \nonumber \\
    +\frac{2}{\phi_0^2V''(\phi_0)}\left(\nabla_\mu \nabla_\nu T-\eta_{\mu \nu} \Box T\right)\ .
    \label{waveqTmunu2}
\end{align}
Hence, the generation of the GWs does not depend only on the presence of perturbed sources of matter but also on the way the energy-momentum tensor varies, as denoted by the presence of second partial derivatives in the equation (\ref{waveqTmunu2}). However, one would expect that for matter sources varying very slowly, the last two terms of the equation (\ref{waveqTmunu2}) might be neglected, recovering the usual case of GR. Nevertheless, in order to keep the analysis as general as possible and study the possible deviations from GR, we are considering that the last two terms might be the same order as the energy-momentum tensor perturbation itself. To solve the equation (\ref{waveqTmunu2}), we define the effective energy-momentum tensor as:
\be
T_{\mu\nu}^{\text{eff}}=\frac{1}{\phi_0}T_{\mu\nu}-\frac{1}{\kappa^2\phi_0^2V''(\phi_0)}\left(\nabla_\mu \nabla_\nu T-\eta_{\mu \nu} \Box T\right)\ ,
\label{effecT}
\ee
Then, the GW equation (\ref{waveqTmunu}) can be rewritten in the standard GR case as:
\be
  \Box \theta_{\mu \nu}=-2\kappa^2 T_{\mu\nu}^{\text{eff}}\ .
  \label{GWeqGRlike}
\ee
This equation can be solved by following the Green function method, leading to the solution:
\be
\theta_{\mu \nu}=-2\kappa^2\int d^4x' G(x-x')T_{\mu\nu}^{\text{eff}}(x')\ ,
\label{solutionGW}
\ee
where the Green function is provided as far as satisfies the equation:
\be
 \Box G(x-x')=\delta^{4}(x-x')\ .
\label{GreenEq}
\ee
The solution of this equation is given by the retarded Green function:
\be
G(x-x')=-\frac{1}{4\pi |\vec{x}-\vec{x}'|}\delta(x_{ret}^{0}-x^{0\prime})\ .
\label{GreenFunction}
\ee
Here $\vec{x}$ are spatial vectors, $x^{0}=t$ and $x_{ret}^{0}=t_{ret}=t- |\vec{x}-\vec{x}'|$ is the retarded time and we are assuming $c=1$. Then, the solution (\ref{solutionGW}) yields:
\be
\theta_{\mu \nu}=4G\int d^3x' \frac{1}{|x-x'|}T_{\mu\nu}^{\text{eff}}(x')\ .
\label{solutionGW2}
\ee
By considering an enough large distance $|\vec{x}|$ from the source, we can approximate $|\vec{x}-\vec{x}'|=r-\vec{x}'\cdot \hat{n}$, where $r=|\vec{x}|$ and $\hat{n}$ is the unit vector pointing to the direction of the GW propagation. Since the volume integral in (\ref{solutionGW2}) is taken above the source, the transverse-traceless (TT) gauge can be applied, such that we are just interested in the transverse-traceless part which can be obtained by projecting the spatial part of $\theta_{\mu\nu}$ as follows:
\be
\theta_{i j}^{TT}=\Lambda_{ijkl}\theta^{k l}\ ,
\ee
where $\theta_{k l}$ is the spatial part of $\theta_{\mu \nu}$, whose indexes are upper and lower by the flat metric $\eta_{ij}$, whereas the tensor $\Lambda_{ijkl}$ is given by:
\be
\Lambda_{ijkl}(\hat{n})=P_{ik}P_{jl}-\frac{1}{2}P_{ij}P_{kl}\ , \quad P_{ij}=\delta_{ij}-n_{i}n_{j}\ .
\ee
Then, the integral (\ref{solutionGW2}) leads to:
\be
\theta_{ij}^{TT}=\frac{4G}{r}\Lambda_{ijkl}\int d^3x' T^{kl}_{\text{eff}}(t-r+\vec{x}'\cdot\hat{n},x')\ .
\label{solutionGWTT}
\ee
Note that inside the source, the TT gauge can not be applied and the following calculations will not be valid what might imply important differences, as pointed at the end of the paper. Nevertheless, the aim now is to perform a multipole expansion and show the dependence of the quadrupole moment for this class of theories. To do so, let's expand the energy-momentum tensor. For  enough small velocities inside the source, we can make the following expansion:
\begin{align}
T_{kl}^{\text{eff}}(t-\frac{r}{c}+\frac{\vec{x}'\cdot\hat{n}}{c},x')=T_{kl}^{\text{eff}}(t-\frac{r}{c},x')\nonumber \\
+\frac{\partial T_{kl}^{\text{eff}}}{\partial t}\frac{(\vec{x}'\cdot\hat{n})}{c}
+\frac{\partial^2 T_{kl}^{\text{eff}}}{\partial t^2}\frac{(\vec{x}'\cdot\hat{n})(\vec{x}'\cdot\hat{n})}{2c^2}+\mathcal{O}\left(\frac{1}{c^3}\right)\ ,
\label{expansionTeff}
\end{align}
where we have recovered the speed of light to show that the expansion is inversely proportional to powers of $c$. Hence, the expression (\ref{solutionGWTT}) becomes:
\begin{align}
\theta_{ij}^{TT}=\frac{4G}{rc^4}\Lambda_{ijkl}\bigg(S^{kl}+\frac{1}{c}n_{m}\dot{S}^{klm}+ \nonumber \\
\frac{1}{2c^2}n_{m}n_{p}\ddot{S}^{klmp}+...\bigg)\ .
\label{solutionGWTT2}
\end{align}
Here we have defined the tensors $S^{kl..}$ as follows:
\be
S^{kl}=\int d^3x' T_{\text{eff}}^{kl}\ , \quad S^{klm}=\int d^3x' T_{\text{eff}}^{kl} x^{m\prime}\ .
\label{TensorsS}
\ee
Let us just focus on the leading term in (\ref{solutionGWTT2}). To write such a term in a more familiar way, we use the fact that:
\be
\partial_{\mu}T^{\mu\nu}_{\text{eff}}=0 \ .
\label{consTmunu}
\ee
Note also that this implies automatically the conservation of the energy-momentum tensor $\partial_{\mu}T^{\mu\nu}=0$, since the two remaining terms in (\ref{effecT}) cancel each other when taking the divergence. Then, the leading term in (\ref{TensorsS}) yield:
\bea
S^{kl}&=&\int d^3x' T_{\text{eff}}^{km}\delta_{m}^{l} \nonumber \\
&=&\int d^3x' \frac{\partial}{\partial x^{m\prime}}\left(T_{\text{eff}}^{km}x^{l\prime}\right)-\int d^3x' \frac{\partial T_{\text{eff}}^{km}}{\partial x^{m\prime}}x^{l\prime}= \nonumber\\
&=&-\int d^3x' \frac{\partial T_{\text{eff}}^{km}}{\partial x^{m\prime}}x^{l\prime}=\int d^3x' \dot{T}_{\text{eff}}^{k0}x^{l\prime} \nonumber \\
&=&\int d^3x' \ddot{T}_{\text{eff}}^{00}x^{k\prime}x^{l\prime}\ .
\label{expansionS}
\eea
Here we have used the continuity equation (\ref{consTmunu}) and the volume of integration is taken over the source, such that boundary terms are removed. Moreover, in the last step, the same procedure as previous steps is followed. Then, by taking now the expression for the effective energy-momentum tensor (\ref{effecT}), the expression for the quadrupole moment tensor is finally obtained:
\begin{align}
S^{kl}&=\frac{1}{\phi_0}\frac{\partial^2}{\partial t^2}\int d^3x' \left(T^{00}-\frac{1}{\kappa^2\phi_0V_0^{\prime\prime}}\partial_{j}\partial^{j}T\right)x^{k\prime}x^{l\prime} \nonumber \\
&=\ddot{Q}^{kl}\ ,
\label{Quadrupole1}
\end{align}
where 
\be
Q^{kl}=\frac{1}{\phi_0}\int d^3x' \left(T^{00}-\frac{1}{\kappa^2\phi_0V_0^{\prime\prime}}\partial_{j}\partial^{j}T\right)x^{k\prime}x^{l\prime}\ 
\label{Quadrupole2}
\ee
is the quadrupole moment for Palatini $f(R)$ gravity. The first term in the r.h.s. of (\ref{Quadrupole2}) corresponds to the usual quadrupole of GR but with an effective gravitational constant given by $G_{eff}=\frac{G}{\phi_0}$ while the second one is an extra contribution that arises in this class of gravitational theories. However, this second term turns out to vanish in the transverse-traceless gauge, when it is substituted in equation \eqref{solutionGWTT2}. Let us show this fact by integrating this term by parts:
\bea
    \int d^3x' \partial_{j}\partial^{j}Tx^{k\prime}x^{l\prime}&=&\int d^3x' \partial_{j}\left(\partial^{j}Tx^{k\prime}x^{l\prime}\right) \nonumber \\-\int d^3x' \partial^{j}\left(T\partial_{j}\left(x^{k\prime}x^{l\prime}\right)\right)&+&\int d^3x' T\partial_{j}\partial^{j}\left(x^{k\prime}x^{l\prime}\right)\ , \nonumber\\
    &=&2\int d^3x' T\delta^{kl}\ ,
    \label{integratebyparts}
\eea
where again the boundary terms are zero because the volume of integration is taken over the source. In this way, the second term in the r.h.s. of (\ref{Quadrupole2}) is proportional to $\delta^{kl}$ and vanishes when computing \eqref{solutionGWTT2}, since $\Lambda_{ijkl}\delta^{kl}=0$. Therefore, the functional form of the quadrupole moment is equivalent to GR result when taking the transverse-traceless projection up to a rescaling of the gravitational constant.

At this point, one can continue the analysis by computing higher order terms in the multipolar expansion.
In fact, by following a similar procedure as in \eqref{expansionS}, an expression for $\dot{S}^{klm}$ in terms of $\dddot{T}_{\text{eff}}^{00}$ and $\ddot{T}_{\text{eff}}^{k0}$ can be obtained. With this purpose, we use the definition in \eqref{TensorsS} for the octupolar term and the fact that the energy-momentum tensor is conserved \eqref{consTmunu}. However, in this case due to the cyclic behaviour of the integration by parts, we can  obtain the following two implicit equations:
\bea
    \dot{S}^{klm}+\dot{S}^{kml}&=&-\int d^3x' \ddot{T}_{\text{eff}}^{k0}x^{l\prime}x^{m\prime}\ ,    \label{s1}\\
    \dot{S}^{klm} + \dot{S}^{lmk}+\dot{S}^{mkl}&=&\frac{1}{2}\int d^3x' \dddot{T}_{\text{eff}}^{00}x^{k\prime}x^{l\prime}x^{m\prime}\ .
    \label{s2}
\eea
By combining both equations, \eqref{s1} and \eqref{s2}, an explicit expression for $\dot{S}^{klm}$ is obtained:
\begin{align}
    \dot{S}^{klm}=\frac{1}{2}\int d^3x' \dddot{T}_{\text{eff}}^{00}x^{k\prime}x^{l\prime}x^{m\prime}+\int d^3x' \ddot{T}_{\text{eff}}^{m0}x^{l\prime}x^{k\prime}\ .
    \label{octupole}
\end{align}
Here, for convenience, let us split the effective energy-momentum tensor into two terms:
\begin{equation}
    T_{\mu\nu}^{\text{eff}}=T_{\mu\nu}^{\text{GR}}+T_{\mu\nu}^{\text{PA}}\ ,
\end{equation}
where,
\bea
    T_{\mu\nu}^{\text{GR}}&=& \frac{1}{\phi_0}T_{\mu\nu}\ ,     \label{GRT}\\
    T_{\mu\nu}^{\text{PA}}&=&-\frac{1}{\kappa^2\phi_0^2V''(\phi_0)}\left(\nabla_\mu \nabla_\nu T-\eta_{\mu \nu} \Box T\right)\ .
    \label{PAT}
\eea
Note that equation \eqref{GRT} is the usual energy-momentum tensor of GR except for the effective gravitational constant while equation \eqref{PAT} is the contribution of Palatini $f(R)$ theories. With that in mind, we can also split \eqref{octupole} into its GR-like part and Palatini $f(R)$ contribution. The latter reads:
\begin{align}
    \dot{S}_{\text{PA}}^{klm}=\frac{1}{2}\int d^3x' \dddot{T}_{\text{PA}}^{00}x^{k\prime}x^{l\prime}x^{m\prime}+\int d^3x' \ddot{T}_{\text{PA}}^{m0}x^{l\prime}x^{k\prime}\ .
    \label{octupolePA}
\end{align}
Then, by inserting equation \eqref{PAT} into \eqref{octupolePA}, it leads to:
\bea
    \dot{S}_{\text{PA}}^{klm}&=&-\frac{1}{\kappa^2\phi_0^2V''(\phi_0)}\frac{\partial^3}{\partial t^3}\bigg[\frac{1}{2}\int d^3x' \partial_j\partial^j T x^{k\prime}x^{l\prime}x^{m\prime}\nonumber \\
    &+&\int d^3x' \partial^m T x^{l\prime}x^{k\prime} \bigg]\ .
    \label{octupolePA2}
\eea
Hence, analogously to \eqref{integratebyparts} and keeping in mind that boundary terms vanish, the expression for $\dot{S}^{klm}$ is reduced to:
\begin{align}
    \dot{S}_{\text{PA}}^{klm}=-\frac{1}{\kappa^2\phi_0^2V''(\phi_0)}\frac{\partial^3}{\partial t^3}\int d^3x' T  \delta^{kl}x^{\prime m}\ .
\end{align}
Whereas the corresponding contraction with the normal vector $n$ yields:
\begin{align}
    n_m\dot{S}_{\text{PA}}^{klm}=-\frac{1}{3\kappa^2\phi_0^2V''(\phi_0)}\frac{\partial^3}{\partial t^3}\int d^3x' T \delta^{kl}n_m x^{\prime m}\ .
\end{align}
Recall that the lambda tensor vanishes when contracted with the Kronecker delta, i.e.  $\Lambda_{ijkl}\delta^{kl}=0$, this implies that the Palatini $f(R)$ contribution to the octupolar term to equation \eqref{solutionGWTT2} vanishes as well. Hence, the octupole order turns out the same as in GR:
\be
    \dot{S}^{klm}=\frac{1}{2\phi_0}\int d^3x' \dddot{T}^{00}x^{k\prime}x^{l\prime}x^{m\prime}+\frac{1}{\phi_0}\int d^3x' \ddot{T}^{m0}x^{l\prime}x^{k\prime}\ .
    \label{octupoleFinal}
\ee
One might go further in the multipole expansion, but same results will be obtained since the transverse-traceless projections over the Palatini contribution to the energy-momentum tensor vanish. Hence, in the weak-field limit, the multipolar expansion in Palatini $f(R)$ gravity coincide functionally with the GR predictions. In this sense, if one just keeps the lowest order of the expansion (quadrupole order),  the expression for the gravitational wave (\ref{solutionGWTT2}) reads:
\be
\left[\theta_{ij}^{TT}\right]_{\text{quad}}=\frac{4G}{rc^4}\Lambda_{ijkl}\ddot{Q}^{kl}\ ,
\label{solutionGWTT3}
\ee
where the quadrupole moment (\ref{Quadrupole1}) reduces to:
\begin{align}
Q^{kl}&=\frac{1}{\phi_0}\int d^3x' T^{00} x^{k\prime}x^{l\prime}\ .
\label{QuadrupoleGR}
\end{align}
While the energy radiated by the GWs emission is given by:
\be
\left[\frac{d E}{dt} \right]_{\text{quad}}\propto \langle \ddot{Q}^{kl} \ddot{Q}^{kl}\rangle\ ,
\ee
where $\langle \rangle$ define a temporal average. Then, despite the radiated energy owns the same dependence over the 00-component of the energy-momentum tensor as in GR, it is rescaled by a factor $1/\phi_0^2$, such that one might conclude that GWs observations might constrain the value of $\phi_0$ as far as the time variation of the energy density is known. Nevertheless, searching for these probes on actual GW events is not an easy task. Current observations are using a variety of methods (see, for example \cite{LIGOScientific:2019fpa, LIGOScientific:2020tif, LIGOScientific:2021sio, Krishnendu:2021fga,
Krishnendu:2021fga}) to test GR and, so far, they are being successful. In our case, we require a direct measurement of the radiated energy before the merger, during the inspiral phase, as our multipolar expansion of the energy-momentum tensor is only valid in this regime. However, up to our knowledge, measurements of the radiated energy on the GW events are usually carried out by computing the difference between the masses before and after the merger (see, for example \cite{LIGOScientific:2016aoc,LIGOScientific:2016sjg,LIGOScientific:2017bnn}), where the masses are computed by considering GR. Therefore, in order to test our results and constrain the value of $\phi_0$, an alternative way to measure the radiated energy within the inspiral phase is necessary. In addition, since $\phi_0$ and the gravitational constant arise together in (\ref{solutionGWTT3}), one also requires a way to measure both separately, otherwise any gravitational system would be coupled to the effective gravitational constant $G_{eff}=G/\phi_0$ and one might not disentangle both values.

%%%%%%%%%%%%%%%%%%%%%%%%%%%%%%%%%%%%%%
\section{Conclusions}
\label{conclusions}
%%%%%%%%%%%%%%%%%%%%%%%%%%%%%%%%%%%%%%

Along this paper, the analysis of GWs propagation and emission is carried out for $f(R)$ theories within the Palatini formalism. To do so, firstly the field equations are expressed in a scalar-tensor-like form, which lead to the presence of a non-dynamical scalar field but sourced by the presence of matter, according to the scalar field equation itself. Then, the full set of field equations are linearised at first order in perturbations of the spacetime metric and the scalar field over an arbitrary background.  To analyse the weak-field regime, we focus on studying first order tensor perturbations of the metric in vacuum and also with the presence of weak  sources. For the former case, the usual wave equation of GR is recovered, such that GWs propagation within the Palatini formalism obey the same rules as in GR with respect to polarisations and the speed of propagation. \\

When assuming the presence of weak sources and analysing the emission of GWs through retarded Green functions, despite that one finds that the GWs equation is sourced by an effective energy-momentum tensor that depends on derivatives of the trace of the energy-momentum tensor, the quadrupole moment just depends on time variations of the energy density, as provided by the 00-component of the energy-momentum tensor. The octupole moment is neither sourced by the Palatini contribution. Higher orders on the multipole expansion will lead to the same GR results. The reason behind lies on the fact that the Palatini contribution to the effective energy-momentum tensor becomes null when taking transverse-traceless projections. Hence, the same functional dependence as in GR is recovered except for a rescaling of the gravitational constant by the expected value of the scalar field in vacuum. One should note that this rescaling might have important consequences, since the energy radiated by the GWs emission will differ with respect to GR's predictions unless $\phi_0\sim 1$. Hence, a remarkable point is that these results might constrain the form of the gravitational action, since every $f(R)$ action within the Palatini formalism will lead to different radiated energy as the value of the scalar field in vacuum depends directly on the form of the gravitational action. \\

Moreover, here we have considered just weak sources where perturbations arise over a flat background. One would expect that in strong-field regimes, the emission of GWs will deviate further from GR predictions showing up a stronger dependence on the gravitational action $f(R)$ or in other words, on the potential of the scalar field and not just on the roots of $V(\phi_0)=0$, as it does for the weak-field limit analysed in this paper. Moreover, the preliminar expressions for the quadrupole (\ref{Quadrupole1}) and octupole moments (\ref{octupolePA2}) show an inverse dependence with respect to the effective scalar field mass $m_{eff}^2=V''(\phi_0)$ that suggest that in strong-field regimes, light scalar fields might induces large deviations with respect to GR predictions whereas for an enough massive scalar field, such corrections would be neglected.\\

Hence, our analysis suggest that next steps should lie on analysing perturbations in strong-field regimes for Palatini-like theories, since in order to constrain this type of theories, a complete analysis of the whole coalescing process is required, as pointed out at the end of section \ref{Emission}. In addition, the extension of these analysis to any arbitrary RBG theories will provide a grateful insight on the understanding of theories within the Palatini formalism.

%%%%%%%%%%%%%%%%%%%%%%%%%%%%%%%%%%%%%%
\section*{Acknowledgments}
 This work is supported by the Spanish National Grant PID2020-117301GA-I00 (ADC and DSG) funded by MCIN/AEI/10.13039/501100011033 (``ERDF A way of making Europe" and ``PGC Generaci\'on de Conocimiento"). ADC is also funded by a pre-doctoral contract from the Predoctoral Contracts UVa 2022 co-financed by Banco Santander.

\end{document}